\documentclass[aps,prl,twocolumn,groupedaddress,preprintnumbers]{revtex4}

\usepackage{graphicx}
\usepackage{dcolumn}
\usepackage{bm}
\begin{document}
\title{\bf Triplon Localization Effect in Tl$_{1-x}$K$_x$CuCl$_3$}

\author{Yosuke Shindo}
\affiliation{Department of Physics, Tokyo Institute of Technology, Meguro-ku, Tokyo 152-8551, Japan}

\author{Hidekazu Tanaka}
\affiliation{Research Center for Low Temperature Physics, Tokyo Institute of Technology, Meguro-ku, Tokyo 152-8551, Japan}

\date{\today}

\begin{abstract}
The effect of randomness on field-induced magnetic ordering was investigated through specific heat measurements in Tl$_{1-x}$K$_x$CuCl$_3$ with $x\leq 0.22$. The isostructural parent compounds TlCuCl$_3$ and KCuCl$_3$ are coupled spin dimer systems with a gapped ground state and their field-induced antiferromagnetic ordering is described by the Bose condensation of spin triplets (triplons). Well-defined field-induced phase transitions were observed in Tl$_{1-x}$K$_x$CuCl$_3$. The critical exponent $\phi$ of the phase boundary defined by $T(H) \propto (H-H_{\rm c})^{1/\phi}$ is reevaluated in TlCuCl$_3$ as $\phi=1.67 \pm 0.07$, which is close to $\phi_{\rm BEC} =3/2$ derived from the triplon BEC theory. For $x \neq 0$, the exponent $\phi$ decreases systematically with $x$. The phase boundary observed at low temperatures for $x>0.1$ is almost a linear function of temperature $T$. In the low-field region for $H<H_{\rm c}$, no magnetic ordering is observed in spite of finite susceptibility. These properties are discussed in connection with the Bose glass phase argued by Fisher {\it et al.} [Phys. Rev. B {\bf 40}, 546 (1989)].

\end{abstract}
%\keywords{}

\narrowtext
\maketitle
\label{sec:intro}
Quantum phase transitions in coupled antiferromagnetic spin dimer systems are of current interest. These systems often have a gapped singlet ground state. In a magnetic field, created $S_{z}=+1$ component of spin triplet which has recently been called triplon \cite{Schmidt} can hop to neighboring dimers and interact with one another due to the transverse and longitudinal components of the interdimer interaction. Hence, the system can be represented as a system of interacting triplon bosons \cite{Rice}. When the hopping term is dominant, triplons can undergo Bose-Einstein condensation (BEC) in a magnetic field higher than the critical field $H_{\rm c}$ corresponding to the gap. This leads to field-induced transverse magnetic ordering \cite{Giamarchi,Nikuni,Wessel}. On the other hand, when the repulsive interaction due to antiferromagnetic interdimer interactions is dominant, triplons can form a superlattice accompanied by a magnetization plateau as observed in SrCu$_2$(BO$_3$)$_2$ \cite{Kodama}. 

TlCuCl$_3$ and KCuCl$_3$ have the same monoclinic crystal structure composed of chemical dimer Cu$_2$Cl$_6$ \cite{Willett,Takatsu}, in which Cu$^{2+}$ ions have spin-$\frac{1}{2}$; their magnetic ground states are spin singlets with excitation gaps $\Delta /k_{\rm B}$ of 7.5 K and 31 K \cite{Shiramura,Oosawa1,Oosawa2}, respectively. The gaps originate from the strong antiferromagnetic exchange interaction between spins in the chemical dimer \cite{Kato,Cavadini1,Cavadini2,Oosawa3}. The neighboring spin dimers couple antiferromagnetically in three dimensions. Field-induced magnetic ordering in TlCuCl$_3$ has been extensively studied by various techniques \cite{Oosawa1,Oosawa4,Tanaka,Sherman}. The results obtained were in accordance with the triplon BEC model \cite{Nikuni}. Magnetic excitations in magnetic fields were investigated by neutron inelastic scattering \cite{Rueegg}, and the results were clearly explained by a theory using the bond operator method \cite{Matsumoto1,Matsumoto2}.

Fisher {\it et al.} \cite{Fisher} theoretically discussed the behavior of lattice bosons in random potential, and argued that a new {\it Bose glass} phase exists at $T=0$ in addition to superfluid and Mott insulating phases. In the Bose glass phase, bosons are localized due to randomness, but there is no gap, so that the compressibility is finite. Fisher {\it et al.} showed that the superfluid transition occurs only from the Bose glass phase, and that near $T=0$, the transition temperature $T_{\rm c}$ is expressed as 
\begin{eqnarray}
T_{\rm c} \sim [\rho_{\rm s}(0)]^x,\ \ \ \ \ \rho_{\rm s}(0) \sim (\rho - \rho_{\rm c})^{\zeta} , 
\end{eqnarray}
where $\rho_{\rm c}$ is the critical density at which the superfluid transition occurs at $T=0$, $\rho_{\rm s}(0)$ is the superfluid density at $T=0$ and exponents $x$ and $\zeta$ are respectively $x=\frac{3}{4}$ and $\zeta\geq \frac{8}{3}$ for three dimensions. The critical behavior is different from that of the boson system without randomness, for which the exponents are given by $x=\frac{2}{3}$ and $\zeta =1$. 

In the present spin system, the superfluid and Mott insulating phases are translated as the field-induced ordered phase and the gapped singlet state or magnetization plateau state, respectively. The strong intradimer interaction $J$ corresponds to the local potential of triplons \cite{Nikuni}. Since $J/k_{\rm B}=65.9$ K and 50.4 K for TlCuCl$_3$ and KCuCl$_3$ \cite{Cavadini1,Cavadini2,Oosawa3}, respectively, we can expect that the partial K$^+$ ion substitution for Tl$^+$ ions produces random on-site potential. Thus, the present system seems suitable for studying boson localization. Then we performed specific heat measurements to investigate the phase transitions in Tl$_{1-x}$K$_x$CuCl$_3$. In this paper, we report the results.

Single crystals of Tl$_{1-x}$K$_x$CuCl$_3$ were grown from a melt by the Bridgman method. The details of preparation were reported in reference \cite{Oosawa5}. Potassium concentration $x$ was determined by emission spectrochemical analysis. Specific heat measurements were performed for samples with $x=0$, 0.055, 0.13 and 0.22. The samples were cut into pieces of 5$\sim$10 mg. We treated samples in a glove box filled with dry nitrogen to reduce the amount of hydrate phase on the sample surface. The specific heats were measured down to 0.45 K in magnetic fields up to 9 T using a Physical Property Measurement System (Quantum Design PPMS) by the relaxation method. The magnetic field was applied along the $b$-axis.

Figure 1 shows the temperature dependence of the total specific heat $C$ in Tl$_{1-x}$K$_x$CuCl$_3$ with $x=0.13$ measured at various magnetic fields. For $H=0$ T, no anomaly indicative of magnetic ordering is observed down to 0.45 K. For $H > 5$ T, a small cusplike anomaly due to magnetic ordering is observed. Arrows indicate transition temperatures. The anomaly at around the ordering temperature $T_{\rm N}$ is small, as observed in TlCuCl$_3$ \cite{Oosawa4}. To determine $T_{\rm N}$ more definitely, we plotted the difference between the specific heat $C(H)$ for a magnetic field $H$ and $C(0)$ for $H=0$, and then the $\lambda$-like anomaly due to the phase transition was clearly observed in $C(H)-C(0)$ versus $T$. The anomaly in the present mixed systems is as sharp as that in the pure system. This is evidence of the good homogeneity of the samples. The specific heat anomaly below 2 K is so small that it is hard to distinguish the ordering temperature. Then, we measured the field dependence of the specific heat at various temperatures. Some examples of the measurements are shown in Fig. 2. The specific heat exhibits a clear cusplike anomaly, to which we assign the transition field $H_{\rm N}(T)$. No hysteresis was observed in the field scan, although small hysteresis was reported in another measurement \cite{Sherman}. Similar measurements were also performed for samples with $x=0$, 0.055 and 0.22. The phase transition points obtained by temperature and field scans are summarized in Fig. 3. Since the phase boundaries for $H{\parallel}b$ and $H{\perp}(1, 0, {\bar 2})$ obtained in the previous magnetization measurements coincide when normalized by the $g$-factor \cite{Oosawa4,Oosawa5}, we infer that the behavior of the phase boundary is independent of external field direction. 

\begin{figure}
\includegraphics[width=6.1cm]{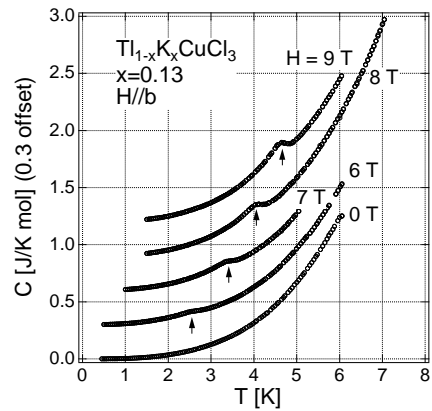}%
\caption{Temperature dependence of the specific heat $C$ in Tl$_{1-x}$K$_x$CuCl$_3$ with $x=0.13$ at various magnetic fields for $H//b$. For clarity, the values of $C$ are shifted upward consecutively by 0.3 J/K mol with increasing external field. Arrows denote the transition temperature $T_{\rm N}(H)$. \label{Fig. 1}}
\end{figure}

We first discuss the phase boundary for pure TlCuCl$_3$. As shown in Figs. 3, the phase boundary can be expressed by the power law 
\begin{eqnarray}
T(H) \propto (H-H_{\rm c})^{1/\phi},
\end{eqnarray}
where $H_{\rm c}$ is the critical field at $T=0$. The best fit is obtained with $H_{\rm c}=5.4 \pm 0.1$ T and $\phi =1.89 \pm 0.06$, using all the data points for $H \leq 9$ T. This value of $\phi$ is somewhat smaller than the previous value of $\phi$ of $2.0\sim2.2$ \cite{Oosawa4,Tanaka}. Phase transition points used for fitting in the previous measurements were concentrated between 2 and 4 K, which leads to the larger critical exponent $\phi$. When we use the data for $T \leq 3$ K or $H - H_{\rm c} \leq 1$ T, we obtain a smaller $\phi$ of $1.67 \pm 0.07$, but this is close to $\phi_{\rm BEC}=3/2$ derived from the triplon BEC theory based on the HF approximation \cite{Nikuni,Fisher}. Setting $x=\frac{2}{3}$ and $\zeta =1$ in eq. (1), we obtain $\phi_{\rm BEC}=(x\zeta)^{-1}=3/2$, because magnetization is proportional to the density of triplons $\rho$, and $\rho - \rho_{\rm c} \sim H-H_{\rm c}$ at $T=0$. The experimental exponent $\phi=1.67$ supports the BEC description of field-induced magnetic ordering in TlCuCl$_3$.

Recently, the deviation of the exponent $\phi$ toward larger value from $\phi_{\rm BEC}=3/2$ was theoretically discussed \cite{Sherman,Nohadani}. Sherman {\it et al.} \cite{Sherman} demonstrated that the larger experimental exponent $\phi$ can be attributed to the relativistic dispersion of the form $\epsilon(k)=\sqrt{\Delta^2+Ak^2}$, which gives better description of the observed dispersion. On the other hand, Nohadani {\it et al.} \cite{Nohadani} showed numerically that the exponent $\phi$ is independent of the interaction parameters. They ascribed the deviation to the temperature-driven renormalization of the triplon dispersion. Both theories predict that when the fitting range ($H-H_{\rm c}$) is reduced, the value of $\phi$ becomes smaller and converges to $\phi_{\rm BEC} = 3/2$, as observed in the present measurements. 

For $x \neq 0$, the phase boundary for $H > 6$ T shifts to the low-temperature side with increasing $x$ (see Fig. 3), {\it i.e.}, transition temperature $T_{\rm N}$ decreases. This behavior can be attributed to the localization effect due to randomness, which prevents triplons from forming coherent state, Bose condensation. The notable feature of the phase boundary is that the exponent $\phi$ obtained with all the data points for $H \leq 9$ T decreases systematically with increasing $x$, {\it i.e.} $\phi=1.84, 1.72$ and $1.40$ for $x=0.055, 0.13$ and 0.22, respectively, but the fitting is not well below 2 K. The enlargement of the phase boundaries below 2K are shown in Fig. 4. It is obvious that the phase boundaries for $x\neq 0$ are almost linear in $T$ in contrast with the phase boundary for TlCuCl$_3$ which meets the field axis perpendicularly. We fit eq. (2) to the phase boundaries, taking field range $H-H_{\rm c} \leq 1$ T. The exponent obtained are $\phi=1.46\pm 0.12, 1.18\pm 0.15$ and $1.11\pm 0.15$ for $x=0.055, 0.13$ and 0.22, respectively. These exponents are much smaller than those obtained with all the data points. The value of $\phi$ for $x=0.055$ becomes close to unity, when $H-H_{\rm c} \leq 0.5$ T. These results indicate that randomness produces qualitative change critical behavior around the quantum critical point. 

The present coupled spin dimer system can be mapped onto the interacting boson system \cite{Rice,Giamarchi,Nikuni}. The local potential on the $i$-th dimer $\mu_i$ is expressed by $\mu_i=g\mu_{\rm B}H-J_i$, where $J_i$ is the intradimer on the $i$-th dimer. The intradimer exchange interaction was evaluated, through neutron inelastic scattering experiments, to be $J/k_{\rm B}=65.9$ K and 50.4 K for TlCuCl$_3$ and KCuCl$_3$, respectively, \cite{Cavadini1,Cavadini2,Oosawa3}. Since these two intradimer interactions are different, the partial K$^+$ ion substitution for Tl$^+$ ions produces the random local potential $\mu_i$ of the triplon. The hopping amplitude $t_{ij}$ and the intersite interaction $U_{ij}$ are given by the transverse and longitudinal components of the interdimer exchange interaction $J_{ij}$. Consequently, the partial K$^+$ ion substitution also produces randomness in $t_{ij}$ and $U_{ij}$. Therefore, our system is not necessarily equivalent to the lattice boson system discussed by Fisher {\it et al.} \cite{Fisher}, where the randomness is introduced only in the local potential and the soft core with on-site interaction is assumed. However, since the problem of the instability of gapped ground state against randomness in quantum spin systems is closely analogous to the particle systems as discussed by Totsuka \cite{Totsuka}, we expect that main results for the lattice boson system discussed by Fisher {\it et al.} are applicable to the present Tl$_{1-x}$K$_x$CuCl$_3$ system.
 
 \begin{figure}
\includegraphics[width=6.1cm]{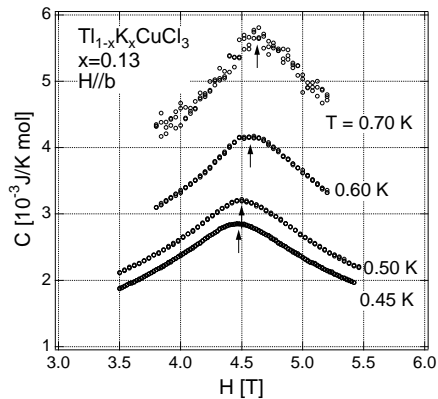}%
\caption{Field dependence of the specific heat $C$ in Tl$_{1-x}$K$_x$CuCl$_3$ with $x=0.13$ at various temperatures for $H//b$. Arrows denote the transition field $H_{\rm N}(T)$. \label{Fig. 2}}
\end{figure}
\begin{figure}
\includegraphics[width=6.1cm]{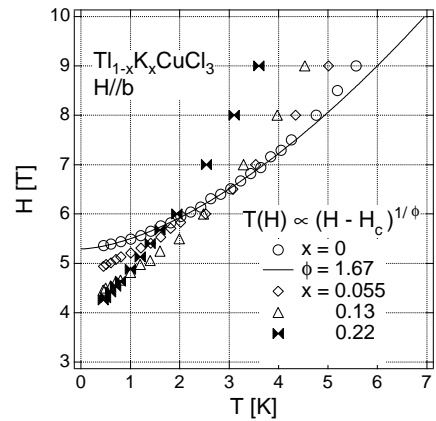}%
\caption{Magnetic field versus temperature diagram obtained for Tl$_{1-x}$K$_x$CuCl$_3$ with various potassium concentrations $x$ for $H//b$. The solid line denotes the fit by eq. (2) with $H_{\rm c}=5.3 \pm 0.1$ T and ${\phi}=1.67$ for TlCuCl$_3$. \label{Fig. 3}}
\end{figure}
\begin{figure}[!]
\includegraphics[width=6.1cm]{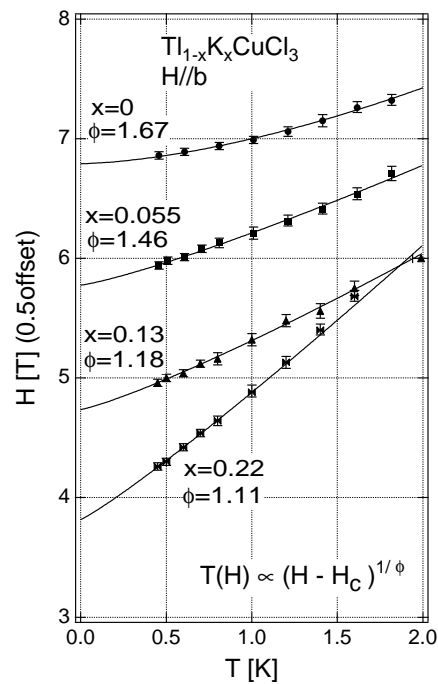}%
\caption{Enlargement of the phase boundaries below 2 K. For clarity, the values of the external field $H$ are shifted upward consecutively by 0.5 T with decreasing potassium concentration $x$. Solid lines denote the fits by eq. (2) with exponents $\phi$ shown in Figure. \label{Fig. 4}}
\end{figure}
 
Oosawa and one of the authors \cite{Oosawa5} reported the results of magnetization measurements on Tl$_{1-x}$K$_x$CuCl$_3$ for $x<0.36$. The main results are as follows: The magnetization curve observed at $T=1.8$ K has finite slope for $H < 4$ T. It is evident that the finite magnetization slope is not due to the finite temperature effect, because TlCuCl$_3$ exhibits almost zero magnetization up to the critical field $H_{\rm c}$. The low-temperature susceptibility reaches finite value and has no Curie term, which shows that finite magnetic susceptibility does not arise from impurities or free spins. Therefore, the finite magnetic susceptibility is intrinsic to the ground state of Tl$_{1-x}$K$_x$CuCl$_3$. With increasing magnetic field, magnetization rapidly increases at a critical field indicating field-induced magnetic ordering. On the temperature scan, the magnetization exhibits a cusplike minimum at the transition temperature $T_{\rm N}$ as predicted by the triplon BEC theory based on the Hartree-Fock (HF) approximation \cite{Nikuni}. 

Since the magnetic susceptibility $\chi=\partial M/\partial H$ corresponds to the the compressibility of the lattice boson system $\kappa=\partial \rho/\partial \mu$, the finite magnetic susceptibility below $H_{\rm c}$ for $x\neq 0$ means that the compressibility of the ground state is finite, {\it i.e.}, there is no gap. From phase boundaries shown in Fig. 4, we see that there is a critical field $H_{\rm c}$ of the field-induced magnetic ordering, {\it e.g.,} $H_{\rm c}\approx 4.3$ T for $x=0.13$. In the low-field phase below $H_{\rm c}$, long range magnetic ordering is absent in spite of the finite susceptibility. This implies that triplons are localized due to randomness. The ground state properties below $H_{\rm c}$ is consistent with the characteristics of the Bose glass phase discussed by Fisher {\it et al.} \cite{Fisher}. Thus, we conclude that the ground state for $H < H_{\rm c}$ in Tl$_{1-x}$K$_x$CuCl$_3$ is the Bose glass phase of triplons.

The value of critical field $H_{\rm c}$ decreases with increasing $x$ as shown in Figs. 3 and 4. This behavior is connected with the $x$ dependence of the lowest singlet-triplet excitation energy, which decreases with increasing $x$ and reaches the bottom at about $x\sim 0.2$ \cite{Niedermayer}. The excitation gap in TlCuCl$_3$ is decreased under hydrostatic pressure, and TlCuCl$_3$ undergoes pressure-induced quantum phase transition to the antiferromagnetic state \cite{Tanaka2,Oosawa6}. Since the ion radius of K$^+$ ion is smaller than that of Tl$^+$ ion, substituting K$^+$ ions for a part of Tl$^+$ ions produce not only the exchange randomness, but also the compression of crystal lattice. Thus, the decrease of $H_{\rm c}$ with $x$ should be ascribed to the chemical pressure due to the ion substitution. 

Applying the theory by Fisher {\it et al.} \cite{Fisher}, we can expect that the phase boundary of Tl$_{1-x}$K$_x$CuCl$_3$ with $x\neq 0$ is described by the power law eq. (2) with a exponent $\phi \leq 1/2$ in the vicinity of $T=0$. The phase boundary for $x\neq 0$ should be tangential to the field axis at $T=0$ in contrast with the phase boundary for TlCuCl$_3$ which is perpendicular to the field axis. The critical behavior represented by the small exponent $\phi \leq 1/2$ is not observed in the present measurements down to 0.45 K. As previously mentioned, however, the low-temperature phase boundary for $x\neq 0$ is almost linear in temperature $T$, {\it i.e.}, $\phi\approx 1$. Such small exponent has not been reported for pure system. In all cases reported, $\phi$ is larger than 2 \cite{Honda,Paduan}. Therefore, we conclude that randomness produces the different critical behavior of field-induced magnetic ordering, and suggest that the $T$-linear behavior arises from the crossover from the convex form ($\phi>1$) to the concave form ($\phi<1$) with decreasing temperature. 

	In conclusion, we have presented the results of specific heat measurements performed on Tl$_{1-x}$K$_x$CuCl$_3$ in magnetic fields. Well-defined field-induced magnetic phase transitions are observed in both temperature and field scans. For TlCuCl$_3$, we reevaluated the critical exponent of the phase boundary to be $\phi =1.67 \pm 0.07$ using the data points for $T < 3$ K. This value is close to $\phi_{\rm BEC} =3/2$ derived from the triplon BEC theory. Randomness produces a qualitative change in critical behavior. For $x\neq 0$, the phase boundary observed below 2 K is almost linear in temperature $T$. The ground state for $H < H_{\rm c}$ in Tl$_{1-x}$K$_x$CuCl$_3$ with $x\neq 0$ has finite magnetic susceptibility, no gap and no long range order. These properties agree with those of the Bose-glass discussed by Fisher {\it et al.} \cite{Fisher}. We conclude that the field-induced magnetic phase transition in Tl$_{1-x}$K$_x$CuCl$_3$ with $x\neq 0$ corresponds to the Bose glass-superfluid transition.

\begin{acknowledgments}
The authors would like to thank A. Oosawa, M. Oshikawa, T. Nikuni and S. Okuma for stimulating discussions. This work was supported by the Toray Science Foundation, and a Grant-in-Aid for Scientific Research on Priority Areas and a 21st Century COE Program at
Tokyo Tech "Nanometer-Scale Quantum Physics" by the Ministry of Education, Culture, Sports, Science and Technology of Japan. 
\end{acknowledgments}

\bibliography{basename of .bib file}

\end{document}